\documentclass[twocolumn,DIV=20,letterpaper,9pt]{scrartcl}
\usepackage[toc,page]{appendix}
\usepackage{ifpdf}
\usepackage{listings}
\ifpdf
\usepackage{graphicx}
\else
\usepackage[dvipdfmx]{graphicx}
\fi

\usepackage{url}

\newcommand{\code}[1]{\texttt{\textup{#1}}}

\newcommand{\Charmpp}{Charm\kern-0.05em\texttt{+\kern-0.03em+}}

\begin{document}

\title{GLB: Lifeline-based Global Load Balancing library in X10}
\author{Wei Zhang, Olivier Tardieu, David Grove, Benjamin Herta, Tomio Kamada (Kobe U/{\sc RIKEN}) \\ Vijay Saraswat,  Mikio Takeuchi\\
~\\ (Besides Tomio Kamada, the other authors are at IBM Research) 
}
\date{December 9, 2013}

\maketitle

\begin{abstract}
We present GLB, a programming model and an associated
implementation that can handle a wide range of irregular parallel programming
problems running over large-scale distributed systems.
GLB is applicable both to problems that are easily
load-balanced via static scheduling and to problems that are hard to
statically load balance. GLB hides the intricate synchronizations
(e.g., inter-node communication, initialization and startup, load balancing,
termination and result collection) from the users. GLB internally uses a version
of the lifeline graph based work-stealing algorithm proposed by
Saraswat et al~\cite{glb}. Users of GLB are
simply required to write several pieces of \textbf{sequential code} that
comply with the GLB interface. GLB then schedules and orchestrates
the parallel execution of the code correctly and efficiently at scale.

We have applied GLB to two representative benchmarks: Betweenness
Centrality (BC)  and Unbalanced Tree Search (UTS). Among them, BC can
be statically load-balanced whereas UTS cannot. In either case, GLB
scales well -- achieving nearly linear speedup on different computer
architectures (Power, Blue~Gene/Q, and K) -- up to 16K cores.

\end{abstract}
\section{Introduction}
\label{sec:intro}

\subsection{Why do we need GLB}
Parallel programming is significantly more challenging than sequential programming.  Given the same input, a sequential program always produces the same result. That is not true for parallel programs. Programmers need to ensure that given the same input, a parallel program can produce the same result under any possible interleaving. In addition, programmers also need to design intricate synchronization schemes to balance the workload among computing resources so that the deployed program can achieve good performance. 

Facing these challenges, many programmers desire a programming model that can can hide the synchronization details from them. Many such programming models have been proposed. Among them, MapReduce~\cite{dean-mr} is one of the most widely used. Programmers only need to provide the sequential mapper and reducer function and the MapReduce framework takes care of input partition, scheduling, inter-machine communication and fault tolerance. While effective for many problems, MapReduce is not applicable to highly-irregular workload. One heavily loaded mapper or reducer can severely downgrade the whole system's performance. Work-stealing~\cite{FrigoLeRa98} is among the first works that handle the irregular workload on shared-memory machine. However, many such techniques are not directly applicable to distributed-memory machines, which are the de-facto programming environment for scale-out computing.

We propose GLB, a Global Load Balancing framework, based on lifeline graph work-stealing algorithm~\cite{glb}. GLB can handle highly irregular problems, where workload on each computing node can be widely different and unpredictable. GLB works on distributed-memory system and can deliver linear speedup and perfect scaling up to 16K cores. In addition, GLB provides a number of parameters for users to tune. Programmers can control the number of random/lifeline victims and task granularity to increase computation throughput or decrease work-stealing response latency, without knowing the underlying complicated synchronization scheme.

\subsection{Why do we choose X10}

X10 is a high-performance, high-productivity programming language for
scale out computing being developed at IBM. It is a class-based,
strongly-typed, garbage-collected, object-oriented
language~\cite{x10-concur05,x10}.  To support concurrency and
distribution, X10 uses the Asynchronous Partitioned Global Address
Space programming model (APGAS~\cite{Saraswat:AMP10}). This model
introduces two key concepts -- places and asynchronous tasks -- and a
few mechanisms for coordination ({\tt finish}, {\tt async},
{\tt at}, {\tt atomic}).  With these, APGAS can express both
regular and irregular parallelism, message-passing-style and
active-message-style computations, fork-join and bulk-synchronous
parallelism.  In contrast to hybrid models like MPI+OpenMP, the same
constructs underpin both intra- and inter-place concurrency.

The X10 language is implemented via source-to-source compilation to
either Java ({\em Managed X10}) or C++ ({\em Native X10}) and is
available on a wide range of operating systems and hardware platforms
ranging from laptops, to commodity clusters, to supercomputers.

Using X10 to implement GLB simplified both the internal implementation
of the library and its end-user API.  X10's high-level support for
distribution and concurrency allowed a concise and efficient
specification of the library and enabled rapid prototyping and
experimentation with design choices.  The X10 language's intrinsic
support for distributed computing, specifically for data serialization
between nodes, simplifies the end-user programming task. Users can
simply specify sequential, single-place data structures and the GLB
implementation can rely on X10 language support to efficiently transmit
user-defined data types between across nodes without requiring any
explicit data serialization code to be written by the user.
 
\section{Design principle}

\subsection{What type of problems are GLB applicable to}
GLB is applicable to problems described as following. There is an initial
bag (multiset) of {\em tasks} (may be of size one). A task usually has a
small amount of associated state, but is permitted to access
(immutable) ``reference state''  that can be replicated across all
places. Consequently, tasks are {\em relocatable}: they can be
executed at any place. 

Tasks can be executed. Executing a task may only involve local,
self-contained work (no communication). During execution, a task is
permitted to create zero or more tasks, and produce a result of a given
(pre-specified) type \code{Z}. The user is required to specify a
reduction operator of type \code{Z}. 

The GLB framework is responsible for distributing the generated
collection of tasks across all nodes. Once no more tasks are left to
be executed, the single value of type \code{Z} (obtained by reducing
the results produced by all tasks) is returned.  

Since the execution of each task depends only on the state of the
task (and on immutable external state), the set of results produced on
execution will be the same from one run to another, regardless of
how the tasks are distributed across places. Since the user supplied
reduction operator is assumed to be associative and commutative, the
result of execution of the problem is determinate. Thus GLB is a
determinate application library that requires the user to provide a
few sequential pieces of code and handles concurrency, distribution
and communication automatically. 

The GLB framework is applicable to a wide variety of tasks. For a
simple example, consider the problem of computing the $n$'th Fibonacci
number in the naive recursive way. Here the state of the task can be
summarized by just one number (long), i.e.{} $n$. Execution of the
task yields either a result (for $n < 2$), or two tasks (for $n-1$ and
$n-2$). The results from all tasks need to be reduced through
arithmetic addition. 

All {\em state space search algorithms} from AI fall in the GLB problem domain. Such algorithms are characterized by a space of states, an
initial state, and a generator function which given a state generates
zero or more states. The task is to enumerate all states (perhaps with
a cutoff), apply some function, and combine the results. An example of
such an application is the famous $N$-Queens problem.

\subsection{Execution overview}
\begin{figure}
\centering
\includegraphics[scale=0.3]{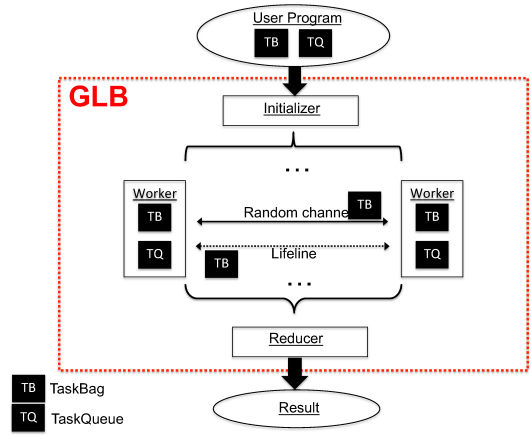}
\caption{GLB overview}
\label{fig:design:glb}
\end{figure}

Figure ~\ref{fig:design:glb} shows the overall flow of GLB. 
For each GLB program, users need to provide two pieces of sequential code \textit{TaskQueue} and \textit{TaskBag}. In addition, users can also provide an \textit{Initialization} method that tells GLB how to initialize the workload at a root place if the workload cannot be statically scheduled among all places. 

\textit{TaskQueue} embodies the sequential computation for this problem and a result reduction function. \textit{TaskBag} embodies the task container's data structure and its split/merge method.    

Given the \textit{TaskQueue} and \textit{TaskBag} implementation, GLB initializes the workload for each \textit{Worker}. Workers are GLB internal computing/load-balancing engines and they are transparent to GLB users. \textit{Workers} process task items by call the \code{process} method provided by \textit{TaskQueue}. \textit{Workers} balance workload using the lifeline based work-stealing algorithm. GLB splits \textit{TaskBag}s from victim \textit{Worker} and merges them to thief \textit{Worker} by calling the \code{split/merge}  methods provided by \textit{TaskBag}. Once all workers finish working, GLB invokes the \code{reducer} function, provided by \textit{TaskQueue}, to return the final result.

In the following sections, Section~\ref{sec:design:user} discusses what are required from users, Section~\ref{sec:design:internals} discusses the GLB internals, Section~\ref{sec:design:uts} discusses how to apply GLB to the UTS problem, Section~\ref{sec:design:bc} discusses how to apply GLB to the BC problem.

\subsection{What do users need to provide}
\label{sec:design:user}
\textbf{TaskQueue} Users need to provide an implementation of following methods:
\begin{enumerate}
\item \code{process(n:Long):Boolean} This method describes the sequential computation of the problem. It iterates over \textit{n} items in its TaskBag and computes each one of them. It processes $n$ items if available and returns true; otherwise it processes all available ($<n$) items and returns false. When GLB sees a Worker's process(n) returns false, it will schedule this Worker to steal from others.
 \item \code{split():TaskBag} Split its TaskBag and returns the split half. This method returns null if the TaskBag is too small to split.  
\item \code{merge(TaskBag)} Merge the incoming TaskBag to its own TaskBag.
Both split and merge method functions are wrapper functions that will call TaskBag's split/merge functions. 
\item \code{getResult()} Returns local result.
\item \code{reduce()} Result reduction function. Applying this function to each Worker's result yields the final result.
\end{enumerate}
\textbf{TaskBag} Users need to provide a custom class extending \textit{TaskBag} and appropriately implementing \code{split} and \code{merge} methods. They provide the functionalities that \textit{TaskQueue}'s \code{split/merge} methods need. GLB provides a default ArrayList-based TaskBag implementation. In this default implementation, split method is implemented by removing half of the elements from the end of the ArrayList and returning this removed half ArrayList; merge method is implemented by adding all the elements from the incoming ArrayList to the local ArrayList.

Once the customized TaskQueue and TaskBag are implemented, it is fairly easy for
users to invoke GLB library by calling GLB's \code{run} method. If the workload cannot be statically scheduled across places, users need to provide an \code{initialize} method so that GLB will initialize this root task at place 0.  

\subsection{GLB internals}
\label{sec:design:internals}
GLB implements the lifeline work-stealing algorithm described in \cite{glb}. We summarize the algorithm here:
\begin{enumerate}
\item
Each Worker repeatedly calls \code{process(n)} method until it is running out of work. Between each process(n) call, Worker probes the network and responds to the stealing requests from other Workers. 
\item
When running out of work, a Worker conducts two rounds of work-stealing. It first chooses at most \textit{w} random victims. If none of the random victims has work to share, then Worker goes to the second round of work-stealing by stealing from its life-line buddies/victims. The topology of life-line communication graph is a $z$-dimensional hyper-cube. Such a topology satisfies these properties: it is a fully connected directed graph (so
work can flow from any vertex to any other vertex) and it has a low
diameter (so latency for work distribution is low) and has a low
degree (so the number of buddies potentially sending work to a dead
vertex is low). If a lifeline buddy cannot satisfy the stealing request due to its lack of work, it will still remember the request and try to satisfy the request when it gets work from others.   
\item
When all Workers run out of work, GLB terminates and applies the reduce function to each Worker's result to yield the final result.
\end{enumerate}

Users can tune the GLB performance by changing these parameters:
\textit{w}, \textit{z}, and \textit{n}. It is more likely to steal
work from a random victim with a larger $w$. It is more likely to
steal work from a lifeline buddy when $z$ is larger. However, when $w$
and $z$ get larger, each worker spends more time probing the network and less time in processing tasks. Alternatively, users can make a Worker spend more time on processing tasks by increasing $n$. The larger $n$ is, the more tasks a Worker needs to finish before it can respond to the stealing requests. However, a large $n$ can sometimes hurt the performance if stealing requests need be responded fast. We will see an extreme case in which even $n=1$ is too large a task-granularity in Section~\ref{sec:design:bc}.

To help users understand and tune GLB program performance, GLB also provides logging functionalities to record (1) how much time each Worker spent on processing and distributing work (2) how many (random/lifeline) stealing requests each Worker sent and received. (3) how many (random/lifeline) stealings each Worker perpetrated. (4) How much workload has each Worker received/sent.

To better illustrate how to use GLB, we show a pedagogical use case via the Fibonacci problem in the Appendix.

\subsection{UTS}
\label{sec:design:uts}
\subsubsection{Problem statement}
The Unbalanced Tree Search benchmark measures the rate of traversal of
a tree generated on the fly using a splittable random number
generator. For this submission, we used a \emph{fixed geometric law}
for the random number generator with branching factor $b_0=4$, seed
$r=19$, and tree depth $d$ varying from 13 to 20 depending on core counts and architecture.

\begin{quote}
The nodes in a geometric tree have a branching factor that follows a
geometric distribution with an expected value that is specified by the  
parameter $b_0 > 1$. The parameter $d$ specifies its maximum depth
cut-off, beyond which the tree is not allowed to grow ... The expected
size of these trees is $(b_0)^d$, but since the geometric distribution
has a long tail, some nodes will have significantly more than $b_0$
children, yielding unbalanced trees. 
\end{quote}

The depth cut-off makes it possible to estimate the size of the trees
and shoot for a target execution time. To be fair, the distance from
the root to a particular node is never used in our benchmark
implementation to predict the size of a subtree. In other words, all
nodes are treated equally, irrespective of the current depth.

Clearly, UTS is a case that static load-balancing does not work.  Assume the sequential implementation of UTS is available (i.e., the code to grow the UTS tree and count the nodes), we now discuss how to apply GLB to it.
\subsubsection{UTS TaskBag and TaskQueue }
\textbf{UTS TaskBag} The internal representation of a UTS tree node is a triple(descriptor, low, high). Descriptor represents the hashed value of the node, low represents the smallest index of un-explored children, high represents the largest index of its un-explored children. The representation of a UTS tree is thus an array of UTS tree nodes.
 A UTS TaskBag is essentially a UTS tree. To split a UTS TaskBag, we evenly split each UTS node $n(d,l,h)$ to two nodes $n1(d,l,h1)$ and $n2(d,h2,h)$, where $h1$ and $h2$ are the middle points of $(l,h)$.  If none of the UTS tree node has more than one child node, then we do not split the tree, as it is cheaper to count the node locally than move it to a remote place and count it there. To merge a UTS TaskBag, we simply concatenate the incoming TaskBag's UTS node array to the local one.

\textbf{UTS TaskQueue} \code{process(n)} method counts at most $n$ UTS tree nodes. \code{reduce()} is a straightforward sum-reduction method on each place's UTS tree node count.

Finally, GLB initializes the workload by initializing the root UTS tree node at Place 0.

\subsection{BC}
\label{sec:design:bc}
\subsubsection{Problem statement}
The Betweenness Centrality benchmark is taken from the SSCA2 (Scalable
Synthetic Compact Application 2) v2.2 benchmark \cite{SSCA2};
specifically, we implement the ``fourth'' kernel in this benchmark,
the computation of {\em betweenness centrality}:

\begin{quotation}
The intent of this kernel is to identify the set of vertices in the
graph with the highest betweenness centrality score. Betweenness
Centrality is a shortest paths enumeration-based centrality
metric, introduced by Freeman (1977). This is done using a betweenness
centrality algorithm that computes this metric for every vertex in the
graph. Let $\sigma_{st}$ denote the number of shortest paths between
vertices $s$ and $t$, and $\sigma_{st}(v)$ the number of those paths
passing through $v$. Betweenness Centrality of a vertex $v$ is defined
as $BC(v) = \Sigma_{s\not=v\not=t \in V} \sigma_{st}(v)/\sigma_{st}$.

The output of this kernel is a betweenness centrality score for each
vertex in the graph and the set of vertices the highest betweenness
centrality score.
\end{quotation}

We implement the Brandes' algorithm described in the benchmark,
augmenting Dijkstra's single-source shortest paths (SSSP) algorithm,
for unweighted graphs. We have implemented the exact variant of the benchmark
(\code{K4approx=SCALE}). 

The solution we implement makes one very strong assumption: that the
graph is ``small'' enough to fit in the memory of a single
place. Since the graph itself is not modified during the execution of
this benchmark, it thus becomes possible to implement this benchmark
by replicating the graph across all places. Now effectively the
computation can be performed in an embarrassingly parallel fashion. The
set of $N$ vertices is statically partitioned among $P$ places; each place is
responsible for performing the computation for the {\em source}
vertices assigned to it (for all $N$ target vertices) and computes its
local \code{betweennessMap}. After this is
done an \code{allReduce} collective performs an \code{AND} summation
    across all local \code{betweennessMaps}. 

Clearly, statically partitioning the work amongst all places in this
way is possible but not ideal. The amount of work associated with one source node
$v$ vs another $w$ could be dramatically different. Consider for
instance a degenerate case: a graph of $N$ vertices, labeled $1\ldots
N$, with an edge $(i,j)$ if $i < j$. Clearly the work associated with
vertex $1$ is much more than the work associated with vertex $N$. 

We next discuss how to dynamically load-balance these
tasks across all places using the GLB framework.

\subsubsection{BC TaskBag and TaskQueue}
\textbf{TaskBag} Each vertex interval is a task item. We use a tuple(low, high) to represent a vertex interval. Each task bag is an array of such tuples. To split a TaskBag, we divide each tuple evenly. To merge a BC taskbag, we simply concatenate the incoming TaskBag's array to the local one.  

\textbf{TaskQueue} \code{process(n)}  method iterates over the taskbag and calculates the first n vertices. 
\textbf{reduce()} method is a simple betweenness-map (a vector) element-wise add function.

This implementation achieves linear speedup and perfect scalability on
small-scale machines (i.e., 256 cores on Power) for smaller
graph. However, it does not yield equally impressive performance on larger machines for larger graphs. After examining the GLB log, we realized that on large scale machines, Workers are less responsive to the stealing requests thus workload cannot be distributed fast. Therefore, we tried to maximize $w$ and $z$ and minimize $n$. (Please refer to Section~\ref{sec:design:internals} for the rationale of turning these parameters). The performance only improved slightly.
We then realized that it took a Worker too long before it responded to the work stealing requests even when its task granularity is \textbf{one} vertex. So we changed the code that computes each vertex to an interruptable state machine. In this way, a Worker can respond to stealing requests without completing one vertex computation. Alternatively, we can help users to minimize the code change by providing a yielding functionality in the GLB library. Users can insert yield points in the their code to increase its probing frequency and responsiveness to stealing requests. We plan to provide such functionality in GLB in the future work.

\section{Performance evaluation}
\subsection{Methodology}
\subsection{Unified code base}
We compare our GLB code to the legacy X10 implementation used to
evalutate X10 performance at Peta-scale~\cite{tardieu-petascale}. To
ensure a fair comparison, we use the same piece of sequential computation code for the legacy code and GLB code. 
\subsection{Platforms}
We evaluated the performance on 3 different architectures: Power~775,
Blue~Gene/Q, and K. We now briefly discuss each architecture and the compiler options we used.

\paragraph{Power-775}
We gathered performance results on a small Power~775 systems with two drawers.
A Power~775 octant (or host or compute node) is composed of a quad-chip module
containing four eight-core 3.84 Ghz Power7 processors, one optical
connect controller chip (codenamed {\em Torrent}), and 128 GB of
memory (in our current configuration). A single octant has a peak performance of 982 GFLOPS; a peak
memory bandwidth of 512 GB/s; and a peak bi-directional interconnect
bandwidth of 192 GB/s. 
operating system image. Eight octants are grouped together into a drawer.

Each octant runs RedHat Enterprise Linux 6.2 and uses IBM's PE MPI for network communication (over PAMI).  

We compiled the benchmark programs using Native X10 version 2.4.0 with \texttt{-NO\_CHECKS}, \texttt{-O} options, and compiled the resulting
C++ files with xlC 12.1 with the \texttt{-qinline} \texttt{-qhot} \texttt{-O3}
\texttt{-q64} \texttt{-qarch=auto} \texttt{-qtune=auto} compiler options.

We allocated 32 places per octant when we use multiple octants.
We used regular 64 KB pages for all the programs.

\paragraph{Blue~Gene/Q}
Our Blue~Gene/Q numbers were gathered on Vesta, a small Blue~Gene/Q
system located at Argonne National Laboratory.
Each compute node of the
Blue~Gene/Q is a 64-bit PowerPC A2 processor with 16 1.6 Ghz compute
cores and 16 GB of DRAM. The compute notes are grouped into drawers
(32 compute nodes per drawer), which are grouped into midplanes (16
compute drawers, 512 compute nodes).  Within the midplane, nodes are
electrically connected in a 5-D torus (4x4x4x4x2). Beyond the
midplane, an optical interconnect is used.  

All benchmarks were compiled using Native X10 version 2.4.0 with \texttt{-NO\_CHECKS}, \texttt{-O} options. The
generated C++ code was then compiled for Blue~Gene/Q using xlC v11.1
with
the \texttt{-O3} \texttt{-qinline} \texttt{-qhot} \texttt{-qtune=qp} \texttt{-qsimd=auto} \texttt{-qarch=qp}
command line arguments.

To maximize the number of X10 Places, all experiments with 16 Places or
more were run using the \texttt{c16} mode, which creates one MPI
process (i.e., one X10 Place) per Blue~Gene/Q compute core.  Thus each
X10 Place has 1 A2 core and 1 GB of DRAM available to it.

\paragraph{K}
The K computer is a supercomputer at RIKEN 
Advanced Institute for Computational Science.
It consists of 82944 compute nodes, and 
each node has one scalar CPU (SPARC64\textsuperscript{\texttrademark}VIIIfx, 8 cores, 128 GFLOPS)
and 16 GB memory.
These compute nodes are connected by Tofu interconnect,
which is 6D mesh/torus network having 10 links (5 GB/s x 2 bandwidth per link) on each node.

Each node runs Linux-based OS, and developers can use MPI-2.1 for network communication.

We compiled the benchmark programs using Native X10 version 2.4.0
with \texttt{-NO\_CHECKS}, \texttt{-O}, and \texttt{-FLATTEN\_EXPRESSIONS} options,
and compiled the resulting C++ files with Fujitsu C/C++ Compiler version K-1.2.0-14 
with \texttt{-Xg} and \texttt{-Kfast} options. 

We allocated 8 places per node when we use multiple compute nodes.
The memory page sizes were set as 32 MB for all the programs.

\subsection{Experimental results}
\subsection{UTS}
In this section, we demonstrate the UTS-GLB performance by showing
three figures, one per each architecture. On each architecture, we
compare the legacy code to the GLB implementation. One thing to note is that the legacy code is a highly tuned parallel implementation of UTS and it won the HPCC2012 performance award~\cite{tardieu-petascale}.
On each figure, the x-axis represents the number of places; the primary y-axis (the left y-axis) represents the number of UTS nodes counted per second; the secondary y-axis (the right y-axis) represents the efficiency, which is calculated by how many nodes counted per second per place. A straight line whose slope is 1 along the primary y-axis indicates a linear speedup and a horizontal line on the secondary y-axis indicates the implementation scales perfectly. In the following discussion, we refer to UTS legacy code by UTS and refer to UTS GLB code by UTS-G.

\begin{figure}
\centering
\includegraphics[width=2.25in]{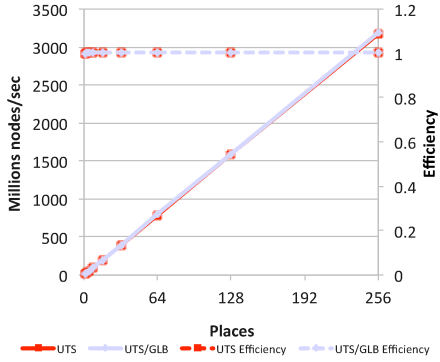}
\caption{UTS/UTS-G Performance Comparison (on Power~775)}
\label{fig:perf:uts-p7ih}
\end{figure}

Figure ~\ref{fig:perf:uts-p7ih} shows the performance comparison for UTS and UTS-G on Power~775 up to 256 cores. UTS and UTS-G both achieve linear speedup and their efficiencies stay at 1, which means they both achieve perfect scaling.  

\begin{figure}
\centering
\includegraphics[width=2.55in]{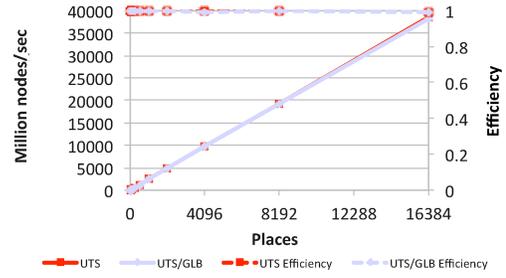}
\caption{UTS/UTS-G Performance Comparison (on Blue~Gene/Q)}
\label{fig:perf:uts-bgq}
\end{figure}
Figure ~\ref{fig:perf:uts-bgq} shows the performance comparison for UTS and UTS-G on Blue~Gene/Q up to 16384 cores. UTS and UTS-G both achieve linear speedup and perfect scaling. 

\begin{figure}
\centering
\includegraphics[scale=0.35]{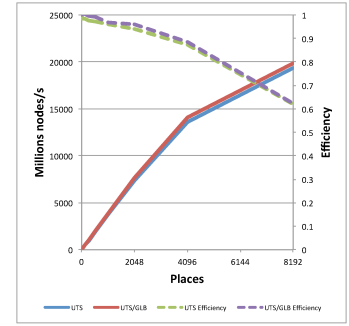}
\caption{UTS/UTS-G Performance Comparison  (on K)}
\label{fig:perf:uts-k}
\end{figure}
Figure ~\ref{fig:perf:uts-k} shows the performance comparison for UTS and UTS-G on Blue~Gene/Q up to 8192 cores. UTS and UTS-G perform well up to 4096 cores (i.e., almost linear speedup and efficiency stays above 0.9). When running on more than 4096 cores, both UTS and UTS-G drop their efficienies to 0.6. We are still investigating why there is a sudden performance drop on K.

\textbf{Summary} UTS-G achieves near linear speedup and perfect scaling on both Power~775 (up to 256 cores) and Blue~Gene/Q (up to 16384 cores). On all three architectures, UTS-G achieves similar (or better) performance compared to UTS, a highly tuned award-winning implementation.

\subsection{BC}
In this section, we compare the BC-GLB performance to the BC legacy
code. Two things to note are: (1) the legacy BC code uses a very
different synchronization scheme from the legacy UTS code. In
comparison, the GLB implementation of UTS and BC share exactly the
same synchronization scheme (i.e., the underlying GLB library). This
demonstrates GLB is applicable to widely different types of
problems. (2) The legacy BC implementation randomizes which vertices
to compute on each place, which effectively reduces the imbalance
among places. As the number of places increases, such imbalance
decreases. When running on more than 1024 places, BC's performance is
only roughly 15\% worse than then optimal performance (i.e., the longest execution time per place is only 15\% more than the average finishing time across all places). Therefore, it is difficult to improve the performance because the room to improve is limited. 

We demonstrate the UTS-GLB performance by showing six figures, two per each architecture. On each architecture, we first compare performance, then we compare the workload distribution among places.   
On each performance comparison figure, the x-axis represents the number of places; the primary y-axis (i.e., the left y-axis) represents the number of edges traversed per second; the secondary y-axis (i.e., the right y-axis) represents its efficiency, which is calculated by number of edges traversed per second per place. A straight line whose slope is 1 along the primary y-axis indicates a linear speedup and a horizontal line along the secondary y-axis indicates the implementation scales perfectly. To understand how effective GLB is at balancing the workload compared to the legacy code, we also plot the workload distribution graphs. On each workload distribution graph, we bar-plot the calculation time on each place and bundle them together. The more even workload distribution graph appears, the more balanced workload is. A rectangular visualization indicates the perfect load-balancing. We also show the mean and standard deviation of the workload distribution on each figure. In the following discussion, we refer to BC legacy code as BC and refer to BC-GLB code as BC-G. 

\begin{figure}
\centering
\includegraphics[width=2.25in]{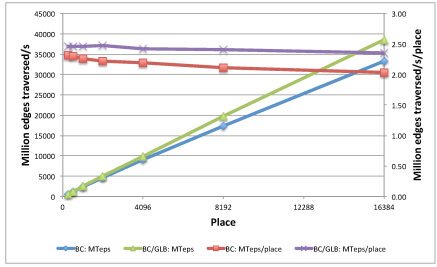}
\caption{BC/BC-G Performance (on Blue~Gene/Q)}
\label{fig:perf:bc-bgq}
\end{figure}
Figure~\ref{fig:perf:bc-bgq} shows the BC performance comparison on Blue~Gene/Q. We can see that GLB implementation achieves near linear speedup and near perfect scaling.

\begin{figure}
\centering
\includegraphics[width=2.25in]{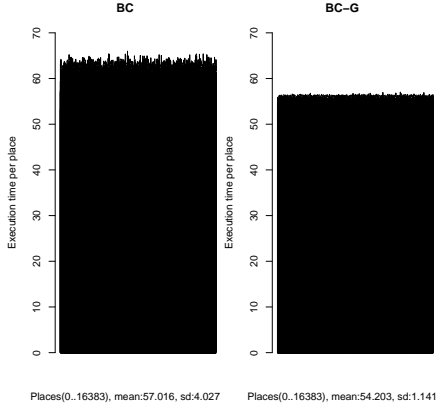}
\caption{BC/BC-G Workload Distribution (on Blue~Gene/Q)}
\label{fig:dist:bc-bgq}
\end{figure}
Figure~\ref{fig:dist:bc-bgq} shows the workload distribution of BC on Blue~Gene/Q. As we can see, BC-G workload distribution is much more even than that of BC, GLB reduces the standard deviation from 4.027 to 1.141. In fact, BC-G finishes calculation in 57.0586 seconds, which is almost equal to the mean of BC computing time accross all places (57.015s). This indicates that BC-G achieves the near perfect load-balancing on Blue~Gene/Q.

\begin{figure}
\centering
\includegraphics[width=2.25in]{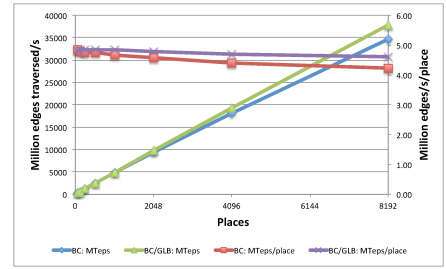}
\caption{BC/BC-G Performance (on K)}
\label{fig:perf:bc-k}
\end{figure}
Figure~\ref{fig:perf:bc-k} shows the BC performance comparison on K. BC-G also achieves near linear speedup and near perfect scaling (efficiency$>$0.95 up to 8192 places).

\begin{figure}
\centering
\includegraphics[width=2.25in]{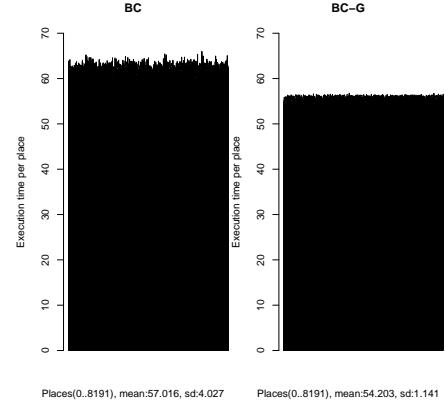}
\caption{BC/BC-G Workload Distribution (on K)}
\label{fig:dist:bc-k}
\end{figure}
Figure~\ref{fig:dist:bc-k} shows the workload distribution of BC on Blue~Gene/Q. As we can see, BC-G workload distribution is much more even than that of BC on K. BC-G finishes calculation in 58.198729 seconds on 8192 places, which is within 1.5\% of the mean of BC computing time across all places (57.016s). This indicates that BC-G also achieves the near perfect load-balancing on K.

\begin{figure}
\centering
\includegraphics[width=2.25in]{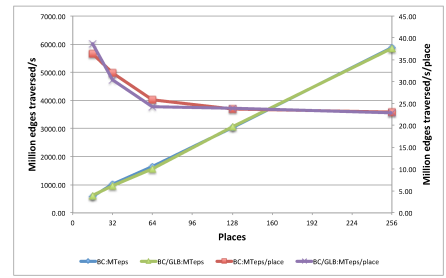}
\caption{BC/BC-G Performance (on Power 775)}
\label{fig:perf:bc-p7ih}
\end{figure}
\begin{figure}
\centering
\includegraphics[width=2.25in]{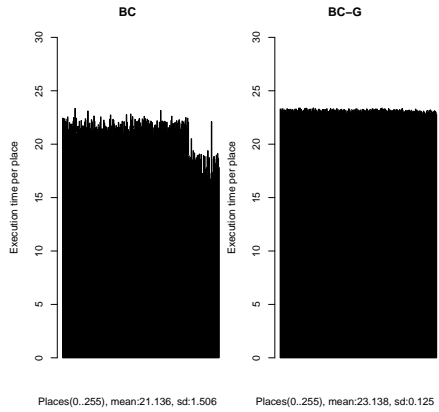}
\caption{BC/BC-G Workload Distribution (on Power 775)}
\label{fig:dist:bc-p7ih}
\end{figure}

Figure~\ref{fig:perf:bc-p7ih} shows the BC performance comparison on
K. Surprisingly, BC-G on Power 775 does not achieve the linear speedup
and perfect scaling as it does on Blue~Gene/Q and on K. That is
because as the number of places increase, the total amount of time to
finish calculating all the vertices also increase even when the graph
stays the same.  Additionally, for the same workload, BC-G spends more
time calculating (on average 5-20\% more) than BC when both run on a
single place. We suspect both problems are due to the overly sensitive
C++ compiler optimizations on Power 755; we are still investigating to
verify this hunch. However, as demonstrated in Figure~\ref{fig:dist:bc-p7ih}, BC-G is very effective at removing the imbalance among workloads, it decreases the workload standard deviation from 58.463 to 1.482.

\textbf{Summary}  BC-G constantly outperforms the BC implementation. On Blue~Gene/Q (up to 16384 cores) and K (up to 8192 cores), BC-G achieves the near-optimal speedup, scalability and load-balancing.

\section{Future work}
We plan to do the following future work: 
(1) Investigate the performance anomaly of UTS-G on K and BC-G on Power 755.
(2) Provide yield points in the GLB library so that users can minimize the changes to the existing sequential code and improve the GLB program's responsiveness to work stealing requests.
(3) Experiment with more benchmarks.
(4) Provide a mechanism to auto-tune GLB parameters (e.g., task granularity, size of random victims/lifeline buddies).

\section{Related work}

There is extensive prior work on dynamic load balancing for
shared-memory systems.  The Cilk~\cite{FrigoLeRa98} system was the
first to provide efficient load balancing for a wide variety of
irregular applications.  Load balancing in Cilk applications is
achieved by a scheduler that follows the \textit{depth-first work,
  breadth-first steal} principle~\cite{blumofe:94}. Cilk's scheduling
strategy, in which each worker maintains its own set of
tasks and steals from other workers when it has nothing to do, is
often referred to as \textit{work-stealing}.  
Following Cilk, many runtimes for shared memory task parallelism
utilize work-stealing schedulers. 
Of these, OpenMP 3.0~\cite{openmp30:08}, Intel's Threading
Building Blocks (TBB)~\cite{reinders:07}, 
Java Fork-Join~\cite{DL2000}, Microsoft's Parallel Patterns
Library (PPL), and Task Parallel Library (TPL) are the most popular.
utilized work-stealing schedulers including X10's
\textit{breadth-first}~\cite{cong:08} and
\textit{compiler-supported}~\cite{TOWH2012} and Guo et
al.'s~\cite{sarkar:09} hybrid model for work stealing.

The techniques for dynamic load balancing in shared-memory
environments are not directly applicable to distributed-memory
machines because of a variety of issues such as network latency and
bandwidth, and termination detection. 
Various bodies of work have addressed the problem of dynamic load balancing on
distributed-memory machines.
Grama et al.~\cite{vipin:94,grama:99} discuss various load balancing 
strategies for distributed parallel searches that are independent of the 
specific search technique.
Blumofe et al.~\cite{blumofe:97} adapt the Cilk work-stealing model to 
distributed shared memory by limiting the scope of the programs to be purely 
functional.
ATLAS~\cite{baldeschwieler:96} and Satin~\cite{nieuwport:01} both use 
hierarchical work-steal to acheive global load balancing.
\Charmpp{}~\cite{sinha:93,kale:93} monitors the execution of its distributed
programs for load imbalance and migrates computation objects to low-load places
to correct the load imbalance.
Global load balancing for message passing environments has been researched 
for specific problems by Batoukov and Sorevik~\cite{batoukov:99}.

UTS, an excellent example of an irregular application, was first described by
Prins et al.~\cite{prins:03}.  It has since been widely used as a
benchmark for dynamic load balancing. 
Olivier and Prins~\cite{olivier:08} provided the first scalable implementation
of UTS on clusters that provided up to 80\% efficiency on 1024 nodes.
To this end, they employed a custom application-level load balancer along with
optimizations such as one-sided communications and novel termination detection
techniques.
Dinan et al.~\cite{dinan:09} provide a framework for
global load balancing, which was used to achieve speedups on UTS on 8196
processors. Global load balancing and termination detection facilities
were provided to express irregular applications.
By reserving one core per compute node on the cluster exclusively for lock and
unlock operations, this framework allowed threads to steal work asynchronously
without disrupting the victim threads. However, the cost paid was a static
allocation (one core out of every eight) for communication. This results in
lower throughput because the thread is not available for user-level
computations.
Saraswat et al.~\cite{glb} introduced lifeline-based global load
balancing and showed 87\% efficiency on 2048 nodes. Later work by the
X10 team demonstrated 98\% parallel efficiency with 55,680 Power7
cores~\cite{tardieu-petascale}.  An implementation of the life-line
algorithm in Co-Array Fortran achieved 58\% efficiency at 8192
nodes~\cite{CAF-HPCC2011}. A more recent UTS code using CAF 2.0 {\tt
  finish} construct achieves a 74\% parallel efficiency on 32,768
Jaguar cores~\cite{Chaoran:2013}.

%

\section*{Acknowledgments}
The code for BC was developed in collaboration with Anju
Kambadur. Earlier versions of the code presented here were worked on
by Sreedhar Kodali, Ganesh Bikshandi, David Cunningham, Pradeep Varma,
Krishna Venkat and other colleagues.  

This material is based upon work supported by the Defense Advanced
Research Projects Agency under its Agreement No.~HR0011-07-9-0002, 
by the Department of Energy and by the Air Force Office of Scientific
Research. 

Some of the results were obtained by using the K computer at the RIKEN
Advanced Institute for Computational Science.  This research was
partially supported by JST, CREST.

This research used resources of the Argonne Leadership Computing
Facility at Argonne National Laboratory, which is supported by the
Office of Science of the U.S. Department of Energy under contract
DE-AC02-06CH11357.

\bibliographystyle{abbrv}
\bibliography{x10}

\begin{thebibliography}{10}

\bibitem{SSCA2}
D.~A. Bader, J.~Feo, J.~Gilbert, J.~Kepner, D.~Koester, E.~Loh, K.~Madduri,
  B.~Mann, and T.~Meuse.
\newblock {HPCS} {S}calable {S}ynthetic {C}ompact {A}pplications \#2: {G}raph
  {A}nalysis.
\newblock
  \url{http://www.graphanalysis.org/benchmark/HPCS-SSCA2_Graph-Theory_v2.2.pdf%
}, 2007.

\bibitem{baldeschwieler:96}
J.~E. Baldeschwieler, R.~D. Blumofe, and E.~A. Brewer.
\newblock Atlas: an infrastructure for global computing.
\newblock In {\em EW 7: Proceedings of the 7th workshop on ACM SIGOPS European
  workshop}, pages 165--172, New York, NY, USA, 1996. ACM.

\bibitem{batoukov:99}
R.~Batoukov and T.~Sorevik.
\newblock {A Generic Parallel Branch and Bound Environment on a Network of
  Workstations}.
\newblock In {\em HiPer '99: Proceedings of High Performance Computing on
  Hewlett-Packard Systems}, pages 474--483, 1999.

\bibitem{blumofe:94}
R.~D. Blumofe and C.~E. Leiserson.
\newblock Scheduling {M}ultithreaded {C}omputations by {W}ork {S}tealing.
\newblock In {\em Proceedings of the 35th Annual Symposium on Foundations of
  Computer Science ({FOCS})}, pages 356--368, 1994.

\bibitem{blumofe:97}
R.~D. Blumofe and P.~A. Lisiecki.
\newblock Adaptive and reliable parallel computing on networks of workstations.
\newblock In {\em ATEC '97: Proceedings of the annual conference on USENIX
  Annual Technical Conference}, pages 10--10, Berkeley, CA, USA, 1997. USENIX
  Association.

\bibitem{cong:08}
G.~Cong, S.~Kodali, S.~Krishnamoorthy, D.~Lea, V.~Saraswat, and T.~Wen.
\newblock Solving {L}arge, {I}rregular {G}raph {P}roblems {U}sing {A}daptive
  {W}ork-{S}tealing.
\newblock In {\em ICPP '08: Proceedings of the 2008 37th International
  Conference on Parallel Processing}, pages 536--545, Washington, DC, USA,
  2008. IEEE Computer Society.

\bibitem{dean-mr}
J.~Dean and S.~Ghemawat.
\newblock Mapreduce: Simplified data processing on large clusters.
\newblock In {\em Proceedings of the 6th Conference on Symposium on Opearting
  Systems Design \& Implementation - Volume 6}, OSDI'04, pages 10--10,
  Berkeley, CA, USA, 2004. USENIX Association.

\bibitem{dinan:09}
J.~Dinan, D.~B. Larkins, P.~Sadayappan, S.~Krishnamoorthy, and J.~Nieplocha.
\newblock Scalable work stealing.
\newblock In {\em SC '09: Proceedings of the Conference on High Performance
  Computing Networking, Storage and Analysis}, pages 1--11, New York, NY, USA,
  2009. ACM.

\bibitem{FrigoLeRa98}
M.~Frigo, C.~E. Leiserson, and K.~H. Randall.
\newblock The implementation of the {C}ilk-5 multithreaded language.
\newblock In {\em Proceedings of the ACM SIGPLAN '98 Conference on Programming
  Language Design and Implementation}, pages 212--223, Montreal, Quebec,
  Canada, June 1998.
\newblock Proceedings published in ACM SIGPLAN Notices, Vol. 33, No. 5, May,
  1998.

\bibitem{grama:99}
A.~Grama and V.~Kumar.
\newblock {State of the Art in Parallel Search Techniques for Discrete
  Optimization Problems}.
\newblock {\em IEEE Trans. on Knowl. and Data Eng.}, 11(1):28--35, 1999.

\bibitem{sarkar:09}
Y.~Guo, R.~Barik, R.~Raman, and V.~Sarkar.
\newblock Work-{F}irst and {H}elp-{F}irst {S}cheduling {P}olicies for
  {A}sync-{F}inish {T}ask {P}arallelism.
\newblock In {\em Proceedings of the 23rd IEEE International Parallel and
  Distributed Processing Symposium}, May 2009.

\bibitem{kale:93}
L.~V. Kal\'{e} and S.~Krishnan.
\newblock {CHARM++}: A portable concurrent object oriented system based on
  {C++}.
\newblock In {\em Proceedings of Object Oriented Programming Systems, Languages
  and Applications, ACM Sigplan Notes}, volume~28, pages 91--108, 1993.

\bibitem{vipin:94}
V.~Kumar, A.~Y. Grama, and N.~R. Vempaty.
\newblock Scalable load balancing techniques for parallel computers.
\newblock {\em J. Parallel Distrib. Comput.}, 22(1):60--79, 1994.

\bibitem{DL2000}
D.~Lea.
\newblock A {J}ava fork/join framework.
\newblock In {\em Proceedings of the ACM 2000 conference on Java Grande}, JAVA
  '00, pages 36--43, New York, NY, USA, 2000. ACM.

\bibitem{CAF-HPCC2011}
J.~Mellor-Crummey, L.~Adhianto, G.~Jin, M.~Krentel, K.~Murthy, W.~Scherer, and
  C.~Yang.
\newblock {Class II Submission to the HPC Challenge Award Competition Coarray
  Fortran 2.0}, Nov. 2011.

\bibitem{olivier:08}
S.~Olivier and J.~Prins.
\newblock Scalable dynamic load balancing using {UPC}.
\newblock In {\em ICPP '08: Proceedings of the 2008 37th International
  Conference on Parallel Processing}, pages 123--131, Washington, DC, USA,
  2008. IEEE Computer Society.

\bibitem{openmp30:08}
{O}penMP {A}rchitecture~{R}eview {B}oard.
\newblock {\em OpenMP Application Program Interface, v3.0}.
\newblock May 2008.

\bibitem{prins:03}
J.~Prins, J.~Huan, B.~Pugh, C.-W. Tseng, and P.~Sadayappan.
\newblock {UPC} {Implementation} of an {Unbalanced} {Tree} {Search}
  {Benchmark}.
\newblock Technical Report 03-034, University of North Carolina at Chapel Hill,
  October 2003.

\bibitem{reinders:07}
J.~Reinders.
\newblock {\em Intel Threading Building Blocks}.
\newblock O'Reilly, 2007.

\bibitem{Saraswat:AMP10}
V.~Saraswat, G.~Almasi, G.~Bikshandi, C.~Cascaval, D.~Cunningham, D.~Grove,
  S.~Kodali, I.~Peshansky, and O.~Tardieu.
\newblock The {A}synchronous {P}artitioned {G}lobal {A}ddress {S}pace {M}odel.
\newblock In {\em AMP'10: Proceedings of The First Workshop on Advances in
  Message Passing}, June 2010.

\bibitem{x10}
V.~Saraswat, B.~Bloom, I.~Peshansky, O.~Tardieu, and D.~Grove.
\newblock The {X10} reference manual, v2.4.
\newblock Sept. 2013.

\bibitem{x10-concur05}
V.~Saraswat and R.~Jagadeesan.
\newblock Concurrent clustered programming.
\newblock In {\em Concur'05}, pages 353--367, 2005.

\bibitem{glb}
V.~A. Saraswat, P.~Kambadur, S.~Kodali, D.~Grove, and S.~Krishnamoorthy.
\newblock Lifeline-based global load balancing.
\newblock In {\em Proceedings of the 16th ACM {S}ymposium on {P}rinciples and
  {P}ractice of {P}arallel {P}rogramming}, PPoPP '11, pages 201--212, New York,
  NY, USA, 2011. ACM.

\bibitem{sinha:93}
A.~B. Sinha and L.~V. Kal\'{e}.
\newblock A load balancing strategy for prioritized execution of tasks.
\newblock In {\em IIPS'93: Proceedings of International Parallel Processing
  Symposium}, pages 230--237, 1993.

\bibitem{tardieu-petascale}
O.~Tardieu, B.~Herta, D.~Cunningham, D.~Grove, P.~Kambadur, V.~A. Saraswat,
  A.~Shinnar, M.~Takeuchi, and M.~Vaziri.
\newblock {X10} and {APGAS} at {P}etascale.
\newblock In {\em Proceedings of the 19th ACM {S}ymposium on {P}rinciples and
  {P}ractice of {P}arallel {P}rogramming}, PPoPP '14. ACM, 2014.

\bibitem{TOWH2012}
O.~Tardieu, H.~Wang, and H.~Lin.
\newblock A work-stealing scheduler for {X}10's task parallelism with
  suspension.
\newblock In {\em Proceedings of the 17th ACM SIGPLAN symposium on Principles
  and Practice of Parallel Programming}, PPoPP '12, pages 267--276, New York,
  NY, USA, 2012. ACM.

\bibitem{nieuwport:01}
R.~V. van Nieuwpoort, T.~Kielmann, and H.~E. Bal.
\newblock Efficient load balancing for wide-area divide-and-conquer
  applications.
\newblock In {\em PPoPP '01: Proceedings of the eighth ACM SIGPLAN symposium on
  Principles and practices of parallel programming}, pages 34--43, New York,
  NY, USA, 2001. ACM.

\bibitem{Chaoran:2013}
C.~Yang, K.~Murthy, and J.~Mellor-Crummey.
\newblock Managing asynchronous operations in coarray fortran 2.0.
\newblock In {\em IEEE 27th International Symposium on Parallel Distributed
  Processing (IPDPS)}, pages 1321--1332, 2013.

\end{thebibliography}

\begin{appendices}
\begin{figure}
  \centering
   \mbox{
\begin{lstlisting}[numbers=left, mathescape,basicstyle=\ttfamily\tiny, firstline=1, lastline=100]
public class FibG(n:Long) {
    class FibTQ implements TaskQueue{
	val bag = new ArrayListTaskBag[Long]();
	var result:Long=0;
	public def getTaskBag()=bag;
	public def getResult()= result;
	public def init(n:Long) {
	    bag.bag().add(n);
	}
	public def process(var n:Long):Boolean {
	    val b = bag.bag();
	    for (var i:Long=0; bag.size() > 0 && i < n; i++) {
		val x = b.removeLast(); // constant time
		if (x < 2) result += x;
		else {
		    b.add(x-1); // constant time
		    b.add(x-2);
		}
	    }
	    return b.size()>0;
	}			
	public def merge(var _tb:glb.TaskBag):void {
	    this.bag.merge(_tb as ArrayListTaskBag[Long]);
	}	
	public def split():glb.TaskBag {
	    return this.bag.split();
	}	
	public def reduce():void {
	    result= Team.WORLD.allreduce(result,Team.ADD); 
	}	
    }
    public def run():Long {
	val init = ()=>{ return new FibTQ(); };
	val glb = new GLB[FibTQ](init, GLBParameters.Default, true);
	val start = ()=>{ (glb.taskQueue()).init(n); };
	glb.run(start);
	PlaceGroup.WORLD.broadcastFlat(()=>{
		(glb.taskQueue()).reduce();
	    });
	return glb.taskQueue().result;
    }
    public static def main(args:Rail[String]) {
	val N = args.size < 1 ? 10 : Long.parseLong(args(0));
	val result = new FibG(N).run();
	Console.OUT.println("fib-glb(" + N + ")" + result);
		
    }
}
\end{lstlisting}}
   \caption{GLB-Fibonacci example}
  \label{fig:design:fib}
\end{figure}
We demonstrate how to use GLB via a Fibonacci example. We show the complete code in Figure~\ref{fig:design:fib}. 
Specifically, We apply GLB to Fibonacci($n$) in the following way: 
\begin{enumerate}
\item \textit{TaskBag} We use the default GLB Arraylist-based Taskbag, as shown at line 3. A task is an integer $i$ whose Fibonacci number should be computed. One can split/merge the TaskBag by calling the \code{merge} and \code{split} methods in the default TaskBag, as shown in Line22 -- Line 27. 
\item \textit{TaskQueue} Each TaskQueue keeps a $result$ of Long type, as shown at line 4.  To process each task item $i$, we first judge if it is less than 2, if so, add $i$ to $result$; if not, remove $i$ from the TaskBag and add $i-1$ and $i-2$ to the TaskBag. When its TaskBag becomes empty, TaskQueue's $result$ is the Fibonacci number of $n$. The \code{process} method is shown in line 10 -- line 21. The reduce function is a simple add function, as shown in line 28  -- line 30. 
\item \textit{Invoke GLB} Since it is difficult to statically schedule tasks among places for the Fibonacci problem, users can invoke GLB by calling the \code{run} method with an \code{initialize} method to use at the root place, as shown in line 35 and line 36. Were it easy to statically load balance, users could call the \code{run} method without providing the initialization method. 
\end{enumerate} 

Note that only these three pieces of code are required from the users and users are oblivious to any synchronization mechanism in X10. 

\end{appendices}
\end{document}